\documentclass[prl,twocolumn,showpacs,amsmath,amssymb,floatfix]{revtex4}

\usepackage{graphicx,times}

\begin{document}

\newcommand{\rem}[1]{}
\newcommand{\LATER}[1]{{}}   
\newcommand{\SAVE}[1]{}
\newcommand{\BRUTAL}[1]{}
\def\OMIT#1 {{}}   

\def\secc#1{{\it #1} ---}
\def\subsecc#1{{\it #1} ---}

\def \beq {\begin{equation}}
\def \eeq {\end{equation}}
\def \bea {\begin{eqnarray}}
\def \eea {\end{eqnarray}}
\def \eqr#1{(\ref{#1})}
\renewcommand{\ss} {{\bf s}}
\newcommand{\nmono} {n_{\rm mono}}
\newcommand{\nx} {n_{\rm x}}
\newcommand{\omegax} {\omega_{\rm x}}
\newcommand{\taux} {\tau_{\rm x}}
\newcommand{\Nflip} {{N_f}}
\newcommand{\Jeff} {{J_{\rm eff}}}
\newcommand{\qq} {{\bf q}}
\newcommand{\PhiSF} {{\Phi_{\rm sf}}}
\newcommand{\TSF} {{t_{\rm sf}}}
\newcommand{\deltaXSF} {{\delta X_{\rm sf}}}
\newcommand{\ellSF} {{\ell_{\rm sf}}}
\newcommand{\ellx} {{\ell_1}}
\newcommand{\deltaPhiSF} {{\delta \Phi_{\rm sf}}}
\newcommand{\ad} {{a_d}}
\newcommand{\apy} {{a_p}}
\newcommand{\rr} {{\bf r}}
\newcommand{\hvec} {{\bf h}}
\newcommand{\hperp} {{\hvec_\perp}}
\newcommand{\rhat} {{\hat{\rr}}}
\newcommand{\Rhat} {{\hat{\RR}}}
\newcommand{\zhat} {{\hat{\zz}}}
\newcommand{\ee} {{\bf e}}
\newcommand{\zz} {{\bf z}}
\newcommand{\RR} {{\bf R}}
\newcommand{\XX} {{\bf X}}
\newcommand{\RRhat}{\hat{\RR}}
\newcommand{\XXhat}{\hat{\XX}}
\newcommand{\comb}{{\rm comb}}
\newcommand{\BB}{{\bf B}}
\newcommand{\Bpar}{{B_\parallel}}
\newcommand{\cD}{{c_D}}
\newcommand{\tobs}{{t_{\rm obs}}}
\newcommand{\tauPhi}{{\tau_\Phi}}
\newcommand{\tauD}{{\tau_0}}
\newcommand{\Eflip}{{E_{\rm flip}}}
\newcommand{\tup}{{t_{\rm up}}}
\newcommand{\Pnuc}{{P_{\rm nuc}}}
\newcommand{\uhat} {{\hat{\bf u}}}
\newcommand{\Bperp} {{\BB_\perp}}
\newcommand{\gbar} {{\bar{g}}}
\newcommand{\Cmono}{C_{\rm mono}}
\def \la {\langle}
\def \ra {\rangle}
\newcommand{\omegaD}{{\omega_D}}


\title{NMR relaxation in spin ice due to diffusing emergent monopoles I}

\author{Christopher L. Henley}
\affiliation{Laboratory of Atomic and Solid State Physics, Cornell University,
Ithaca, New York, 14853-2501}


\begin{abstract}
At low temperatures, spin dynamics in ideal
spin ice is due mainly to dilute, thermally excited 
magnetic monopole excitations.  
I consider how these will affect the dynamics of
a nuclear spin (the same theory applies to muon
spin resonance if implanted muons do not diffuse).
Up to the time scale for nearby monopoles
to be rearranged, a stretched-exponential form of the 
relaxation functions is expected.
I work out the expected exponent in that exponential and
the formulas for the $T_1$ (longitudinal) and $T_2$ (dephasing) 
relaxations,  as a function of the monopole density
(and implicitly the temperature).
\end{abstract}

\pacs{PACS numbers: }

\maketitle

\secc{Introduction}

The dipolar spin ice 
state~\cite{gingras-cancel,spinice-oldrev,monopoles-nature,jaubert-spinice,castelnovo-AnnRev}
occurs at sufficiently low temperatures
in certain highly frustrated pyrochlore structure magnets such as Dy$_2$Ti$_2$O$_7$.
The magnetic ions (here Dy$^{+3}$) sit on a a ``pyrochlore'' magnetic
lattice consisting of corner-sharing tetrahedra, and have a strong
uniaxial anisotropy along their local $\la 111 \ra$ symmetry axis.
The practically degenerate lowest energy states consist of the
(exponentially many) configurations in which every tetrahedron
has two spins pointing inwards and two spins pointing outwards.
The elementary excitations in such a material are defect tetrahedra
with three spins in and one out, or the reverse, which were 
shown to behave as (emergent) magnetic monopole~\cite{monopoles-nature}.
The low-temperature thermodynamics and dynamics of dipolar spin ice
depend on the density and mobility of the monopoles.

Measurements of the specific heat~\cite{spec-ht} and (static) spin correlations 
in the presence of a large field ~\cite{susc-expt},
corroborate the monopole picture.
An analog of the Wien effect in electrolytes was invoked
to infer the effective monopole charge from muon spin relaxation~\cite{bramwell-wien}
but this has been called into question~\cite{dunsiger-muon-critique,blundell-muon-critique}.
The dynamics in magnetic relaxation experiments~\cite{snyder}
shows an activated 
dynamics, but not the activation energy expected from the basic
monopole model~\cite{kycia-Dy-mag}.  Low-temperature nuclear magnetic resonance
(NMR)~\cite{kitagawa-lowT,hiroi-mag-lowT}
or muon spin resonance 
experiments~\cite{blundell-old-muSR,dunsiger-muon-critique,muon-diffusion})
may aid in disentangling this confusion, as they probe {\it local} 
spin dynamics rather than the uniform bulk magnetization.

\OMIT{``Spin ice'' compounds contain moments on
the pyrochlore lattice of corner-sharing tetrahedra,
having an easy local axis anisotropy and frustrated
interactions such that each tetrahedron has two
spins pointing in and two pointing out, in any
(classical) low-energy state.  There are many such configurations,
which are practically degenerate, giving an apparent residual
entropy reminiscent of water ice.}

\BRUTAL{
In this paper I am concerned with 
dipolar-coupled spin ice, in which the elementary excitations
are deconfined defects that behave as magnetic monopoles, and
can diffuse throughout the lattice.  
In a pure spin ice compound 
at low temperatures, diffusing monopoles should be the chief source
of fluctuating magnetic fields, and thus dominate the NMR or muon
spin resonance signals.}

In this Letter, I  will first review the essentials of spin ice and its emergent
monopoles, and then set up the simplest possible model for
NMR relaxation functions,
both the $T_1$ (longitudinal) and $T_2$ (dephasing) kind.
In the presence of disorder that is fixed for the duration of
the measurement --- which includes monopoles if they do not
diffuse too fast ---  the NMR signal is a
superposition of relaxation functions due to
the inequivalent environments of many probe spins,
giving a stretched exponential form.
In the monopole case, the form of the $T_1$ relaxation function 
will be shown to have the same stretched-exponential form as 
the case of fixed magnetic impurities; but the $T_2$ relaxation function
is novel, in that a power of $t^{1/2}$ is replaced by $t^{3/2}$ in
the exponential.
\SAVE{Several regimes (defined by time scale) 
are identified in which distinct behaviors are seen.}
Experiment does find stretched exponentials, but I will show
their parameters are incompatible with a monopole origin.
It is proposed that interstitial ``stuffed'' moments (additional to the
spin ice Hamiltonian) are responsible for the NMR signal 
at very low temperatures.

\secc{Dipolar spin ice and dumbbell model --}

The experiments in Ref.~\cite{kitagawa-lowT}
are on Dy$_2$Ti$_2$O$_7$ in which the magnetic ions 
are rare earth Dy$^{+3}$ (Ho$_2$Ti$_2$O$_7$ is similar).

\LATER{I need to track down
the numerical coefficients -- to write a table of all the parameters
for Dy spin ice and also Ho spin ice.}

Each Dy spin has a strong Ising-like anisotropy 
with easy axis aligned along the $\la 111 \ra$ direction
of its local 3-fold symmetry.
The pyrochlore lattice nearest-neighbor distance 
is $\apy=1/\sqrt{8}$ times the cubic lattice constant.
The tetrahedron centers,
separated by $\ad = \sqrt{3}/4$ times the cubic lattice constant,
constitute the vertices of a diamond lattice,
of which the magnetic (pyrochlore) sites are the bond midpoints.
The dipole interaction of spins at $\rr$ and $\rr'$ is 
     \beq
           D [1-3 \cos\psi \cos \psi']/ |\rr-\rr'|^3 
     \label{eq:Hdipole}
     \eeq
with $D\equiv \mu_0 \mu^2/4\pi \apy^3$,
where $\mu$ is the moment of one spin,
and distances are measured in units of $\apy$.
Also, $\cos\psi$ and $\cos\psi'$ are the angles made by
their respective easy axes with the vector $(\rr-\rr')$.
The effective nearest-neighbor Ising spin interaction is 
$\Jeff= (J+5D)/3 \approx 1.1$K, with a net ferromagnetic
sign 
\SAVE{(even though the exchange interaction $J$ is  antiferromagnetic)}
that is frustrated since the 3-fold axes differ by the tetrahedral angle 
109$^\circ$, enforcing the well-known 
two-in/two-out spin structure in every tetrahedron.

I adopt the ``dumbbell'' approximation for the
spin Hamiltonian of the model system; 
each point dipole is represented by
a pair of opposite magnetic charges on neighboring 
tetrahedron centers.
Then, in any two-in/two-out state, these magnetic charges cancel 
at {\it every} diamond lattice site so that (within this approximation,
and nearly so in dipolar spin ice~\cite{gingras-cancel})
all such states are exactly degenerate.

The elementary excitation is a tetrahedron having 3 in and 1 out-pointing 
spins, or the reverse, which is the ``monopole'' defect (located at the
tetrahedron center).  Each monopole contributes a far field of Coulomb form
$|{\bf B}| = Q/4\pi |\RR|$,
where $Q=\pm 2\mu/a_d$
is its (conserved) magnetic charge,
so they are indeed emergent mangnetic {\it monopoles}~\cite{monopoles-nature}.
Monopole pairs interact with the corresponding Coulomb-like potential, and
even outside the dumbbell approximation each microstate's total energy depends 
almost entirely on the monopole positions and  very little on the configuration 
of spins in 2-in/2-out tetrahedra apart from the monopoles.
where $\mu$ is the electron spin's moment and $a_d$ is the separation
of tetrahedron centers.  

A single spin flip in the ground state
produces a pair of oppositely charged monopoles, which 
after separating can diffuse independently over tetrahedron centers 
or ultimately recombine, much like electrons and holes in a semiconductor.
We let $\Delta_0$ be the cost to create and pull apart a {\it pair} of
monopoles.  Within the dumbbell model, this is set by 
making the nearest-neighbor spin interaction to
match $\Jeff$, by setting 
         $\Delta_0= \frac{4}{3} [ J+ 4 (1+\sqrt{{2}/{3}}) D] 
\approx 6 $K~\cite{jaubert-spinice}.
\SAVE{this is $v_0 Q^2/2$ in the notation of Castelnovo et al,
where $Q=2q_m= 2\mu/a_d$.}
If we ignore Debye screening, 
the thermally excited density of unbound monopoles of {\it both} signs
is thus
   \beq
          \nmono = (\Cmono /v_d) e^{-\Delta_0/2 T}
   \label{eq:nmono}
   \eeq
where $v_d= \sqrt{8} a_p^3$ is the volume per diamond vertex, 
and Pauling's approximation for the entropy gives
$\Cmono \approx 4/3$ for the prefactor.

There are two monopole-dominated
temperature regimes.  Most studies to date 
(including NMR~\cite{kitagawa-takigawa})
focused on moderately low temperatures 1--5 K, 
at which thermally excited monopoles are 
common but dilute enough to be useful degrees of freedom, 
but still dense enough that Debye screening is crucial and 
reduces the effective $\Delta$ in  \eqref{eq:nmono}.
The present study focuses on the very low temperatures 
$T< T^* \approx \Delta/10 \approx 0.6 K$,
at which $\nmono<0.01$ and  screening is negligible.
($T^*$ also appears experimentally as the crossover
between different behaviors~\cite{kitagawa-lowT}.)

{\it Bound pairs} of opposite monopoles are possible with
finite separations. The closest separation is
nearest-neighbor tetrahedra (separation  $a_d$), 
which is equivalent to a flip of   just one spin
(shared by the tetrahedra) relative to the 2-in/2-out background.
This has a cost $\Delta_2$ implying a density of pairs
$\sim \exp(-\Delta_2/T)$.
Although the energy is reduced
by the Coulomb attraction ($\Delta_2< \Delta_0$), 
we lose the factor of 2 in the exponent of \eqref{eq:nmono}
(due to the entropy of deconfinement), so it depends
on details whether bound or unbound monopoles are commoner.
As will be shown below, only the unbound ones make a
novel contribution to the $T_2$ relaxation.
(Following a quench, ``back-to-back'' bound pairs of monopoles  
on adjacent tetrahedra may also persist out of equilibrium,
whenever the intervening spin has
the minority in/out sense.~\cite{castel-back2back}
\SAVE{(The monopoles would have to be
separated farther before they could annihilate.)}

\BRUTAL{
The monopoles may be divided into bound and unbound...
Due to the attractive interaction, the cost of a nearest-neighbor pair
-- which is equivalent to a single flipped spin -- {\it may} be smaller 
than $\Delta/2$.  If so, the single flipped spin is the commonest defect
at low temperatures, but its lifetime is short ($\sim \tau_0$)
compared with that of unbound monopoles.  
The field jumps due to such flips dominate the $T_1$ relaxation 
if their lifetime exceeds $1/\omega_0 \sim 1 \mu$s.
\rem{But, I expect, never dominate the $T_2$ relaxation.}
}

\BRUTAL{
At sufficiently low temperatures, all rearrangements of the 
(electron) spins -- and hence all fluctuations in the field seen 
by probe spins -- {\it must} be due to diffusing monopoles, since
all other excitations cost larger activation energies.
Thus, to represent the spin degrees of freedom,
I further abstract them to a lattice gas of monopoles.
These will be divided into bound pairs with any separation
$|\RR|$ having binding energy $E_b=Q^2/4\pi |\RR|$, so
the density per site is $\sim e^{-(2\Delta-E_b)/T}$, plus
unbound monopoles with density $\sim e^{-\Delta/T}$
All interactions among bound pairs and/or unbound single monopoles
will be neglected.
}

Dy spins are assumed to flip at rate $\tau_0^{-1}$ if the energy 
$\Delta E$ is unchanged or decreases, otherwise $\tau_0^{-1}\exp(-\Delta E/T)$.
The only spin flip with $\Delta E=0$ is by one of the 3 spins in
the majority (in or out) direction on a monopole tetrahedron; 
this transfers the monopole 
to the other tetrahedron sharing the flipped spin.~\cite{FN-hop-anticorrelated} 
Thus the dynamics at low $T$ consists of random-walking monopoles
which have a combined rate to hop of $3\tau_0^{-1}$, plus
the occasional creation or annihilation of a monopole pair.
Based on magnetic relaxation
experiments~\cite{snyder}, it is  {\it believed} that
$\tau_0 = 10^{-8}$ to $10^{-3}$ s and is temperature independent.
(The $\tau_0$ value can be inferred from NMR, as explained further below.)

\secc{NMR model}

To model the zero-field NMR, 
I assume a single probe moment located at the origin,
which for simplicity is imagined to be spin 1/2.~\cite{FN-nuc-moments}
To zero order, the probe sees a base field $\hvec_0$ 
depending on the (frozen) nearby Dy spins, 
the direction of which defines the probe spin's
quantization axis.
Specifically, I have 
in mind an O$^{17}$ nucleus situated at a tetrahedron center:
there, $\hvec_0$ is due mainly to the four surrounding Dy spins
(which satisfy the 2-in/2-out constraint)
and $\hvec_0$ points along a $\la 100\ra$ axis.
\SAVE{(The environment of closest spins with two in/two out always
contributes the {\it same} magnitude $h_0$ at that center --
unless a monopole is sitting on that tetrahedron, which is rare.)}
The base NMR precession frequency is $\omega_0=\gamma h_0$
where $\gamma$ is the gyromagnetic ratio.
(In the case of the Ti$^{47}$ nuclear moment~\cite{FN-NQR}, 
$\omega_0$ instead represents its quadrupolar frequency.)
All nontrivial probe spin dynamics is 
a response to additional, time-varying
perturbing fields $\hvec(t)$ due to distant flipping
spins (presumably Dy: interaction between nuclear moments will
be neglected).
We call these ``flippers'' for short:
they might {\it either} be diffusing monopoles, or fixed spin impurities,
and each flips randomly with a time constant $\taux$.

Within the material, the frozen local magnetic field 
has strong and complicated spatial modulations within each 
unit cell~\cite{field-trip,FN-NMR-split}
However, 
only {\it changes} of $\hvec(t)$ matter for the NMR response:
$T_1$ relaxation is sensitive only to high frequencies --- 
i.e. the flips themselves --- while the $T_2$ relaxation is measured
using the spin-echo technique which cancels the frozen
non-uniformity of the field.
Such changes are given, at separations $R> a_d$,
by the ``Coulomb'' fields of the monopoles,
that is $\pm Q/4\pi R$ in the radial direction.
Actually, since a single monopole hop is the same as 
one spin reversal, the field change $\Delta \hvec$
is just twice the
dipole field due to that spin, which I describe by
a field scale $h_D/R^3$, where the distance $R$ is
measured in units of $\apy$.
We can convert the field scale to a frequency using the 
gyromagnetic ratio, $\omegax = \sqrt{2} \omega_d/R^3$.
\SAVE{$\omegax$ is the typical field due to a dipole
at that distance.}
The factor $\sqrt 2$ is the root-mean-square of the angular
factor in \eqref{eq:Hdipole}.

\SAVE{Thanks to this splitting, we needn't  worry much about 
the diffusion of spin between different nuclei, which is
often important in NMR.}

I will compute both kinds of NMR relaxation functions
$g_1(t)$ and $g_2(t)$. 
To start with, consider the relaxation functions
$g_{01}(t)$ and $g_{02}(t)$ due to a {\it single} flipper 
at distance $|\RR|$ from the probe spin.
To set up the longitudinal or $T_1$ relaxation, say the probe spin
is initially aligned with  its local axis.  As the local field 
direction fluctuates due to the transverse components $\hperp(t)$ of the 
perturbing field, the probe spin follows adiabatically any slow
fluctuations. 
tilts the field axis by a small angle 
$\delta \approx  \Delta\hperp/h_0$;
assuming this changes much faster than a precession period,
the old state is projected onto the new axis, so the
probability for the probe spin to end up with the opposite sense
is $(1-\cos\delta)/2 \approx (\delta/2)^2$.  
The correlation of the probe spin with its original sense thus
decays with time, proportional to the relaxation function 
   \beq
g_1(t) \equiv \Big\langle  (\cos \delta)^\Nflip \Big\rangle \approx
            \exp \Big( - \frac{\Delta\hperp^2}{4 h_0^2} \frac{t}{\taux}\Big),
    \eeq
where the $\Nflip$ is the number of flips in time $t$, its probability 
given by a Poisson
distribution $(t/\taux)^\Nflip \exp(-t/\taux)/\Nflip!$. 

The ``T$_2$'' relaxation
represents dephasing due to the precession of the in-plane angle of tranverse 
polarization due to the fluctuating part of the longitudinal
field, 
$\omega(t) = \gamma (h_\parallel(t)-h_0)$, where $\gamma$ is
the probe spin's gyromagnetic ratio and 
$h_\parallel$
is field component along the probe spina's 
local axis.  
The $T_2$ relaxation is measured using the 
{\it spin-echo} technique: 
the polarization 
is (effectively) flipped by a $\pi$-pulse at a time $t$ 
and then evolved till time $2t$. The relaxation function is thus
    $g_2(2t) = \la \cos(\Phi(2t) \ra$
where 
   \beq
      \Phi(2t) \equiv \int _0^t \omega(t')dt' -\int _t^{2t} \omega(t')\; dt'
   \eeq
is the accumulated phase difference for
precession around the quantization axis.
Thus, the $T_1$ relaxation is dominated by the fluctuations 
at high frequency (compared with $\omega_0$), whereas the $T_2$ 
relaxation is dominated by the slower fluctuations
at frequencies comparable to $1/t$. 
\rem{Lower frequencies or static fields cancel due to the spin-echo flip.}

Now consider specificallythe dephasing relaxation function $g_{02}(\RR;t)$ 
due to a single ``flipper''at distance $|\RR|$.
The change in precession frequency 
at each flip is $\pm 2 \omegax$.
If we know that exactly one spin  flip occurred 
at a random time within the
interval $[0,2t]$, it can be worked out that
$g(2t)= 1-A(2t)$, where $A(2t)\equiv  1-\sin(\omegax t)/(\omegax t)
\approx (\omegax t)^2/6$.  
If the flips in that time are independent, we 
use the Poisson distribution to obtain
    \beq
     g_{02}(\RR;t) = 
           e^{- \frac {2t}{\taux} A(t)}
\approx  e^{-(t/T_{02})^3}
    \label{eq:g2pure}
    \eeq
where 
$T_{02}^{-3} = \frac{1}{3} \omegax^2 \taux^{-1}$ 
The assumption of independent flips is valid for a monopole,
\cite{FN-flips-displacement}.
This is the key difference between the spin
flip statistics of a monopole and those
of a fixed impurity: a conventional Ising impurity
must alternate plus and minus flips since it has only
two states.
\SAVE{Another reason is that the monopole 
ensemble on length scales less than $(Dt)^{1/2}$ 
behaves as ``annealed'' rather than ``quenched'' disorder.}
\BRUTAL{
There is one key difference between a paramagnetic impurity and
a diffusing monopole. Having flipped, an impurity spin can only 
flip back the other way: thus the signs of changes $\Delta \omega$
are strictly anticorrelated. The upshot is that the 
$\exp(- {\rm Const}\; t^3)$ behavior at short times crosses
over to $\exp(-{\rm Const}\; t)$ at $t>\taux$.  
On the other hand, a monopole can keep diffusing away
from its initial position: within our approximation,
each successive change $\Delta \omega$ is an independent
random variable.  The result is that the 
$\exp(- {\rm Const}\; t^3)$ behavior is maintained for
$t>\taux$, which is the regime I assume for the $T_2$ relaxation.
}

In the case of an alternating flipper, 
if $\omegax \taux \gg 1$ then even one flip by the flipper suffices 
to randomize the phase, so $g_{02}(2t)\approx e^{-2t/\taux}$.
On the other hand, if $\omegax \taux \ll 1$ and 
$t/\taux \gg 1$, then $\la \omega(t')\omega(t'')\ra =\omegax^2 e^{-2|t'-t''|/\taux}$
and $\Phi$ has a Gaussian distribution with 
     $\la \Phi(2t)^2\ra= (2t)\omegax^2\taux$, 
implying $g_{02}(2t)\approx e^{-2t/T_{02}}$
with $T_{02}^{-1} = \omegax^2 \taux$.

We can summarize all cases of the single-flipper results by
   \beq
            g_{0i}(\RR;t) \approx e^{-[T/T_{0i}]^{\beta_{0i}}/|\RR|^6}
   \eeq
where
$\beta_{01}=1$, while
$\beta_{02}=1$ for a fixed flipper at $t>\taux$,
but $\beta_{02}=3$ for short times or a diffusing flipper;
the corresponding time constants are 
    \begin{subequations}
    \label{eq:T_0is}
    \begin{eqnarray}
T_{01}^{-1} &=& (\omegaD^2/\omega_0)^2  \taux^{-1}; \\
T_{02}^{-1} &=& \omegaD^2 \taux  \quad \text{ fixed, $t \gg \taux$}; \\
T_{02}^{-3} &=& (\omegaD)^2/\taux  \quad \text{ diffusing or $t \ll \taux$}.
    \end{eqnarray}
    \end{subequations}
and $\omegax \taux \sim \omegaD \taux/|\RR|^3 \ll 1 $ is assumed.`

\secc{Random environments:
stretching exponentials due to inhomogeneity}

When different probe moments have different environments
during the measurement time, the observed signal is 
an average of the relaxation function over these environments 
and is likely to acquire a stretched exponential form.  
This fact is familiar (since the 1960s)
in the case that the field fluctuations
are due to fixed paramagnetic impurities. The same thing happens
if the fluctuations are due to diffusing monopoles whose
displacement during the measurement time is small 
compared to their distance $|\RR|$ from the probe spin.

\BRUTAL{
It has just been shown that the relaxation function of a 
single probe moment should have an 
exponential or stretched-exponential form with 
$\beta_{0i}=1$ or 3.
Smaller stretching exponents $\beta_i$  get produced by  
superposing signals from probe moments that have
different relaxation times.
This behavior has been known for 50 years, for the
case of {\it fixed} paramagnetic impurities.
There are no fixed impurities in our basic spin ice model,
but the distant monopoles play that role, so long
as their displacement, during the experimental time $t$,
is small compared to their distance $R$ from the probe moment.}

I next work through the universal form for the relaxation
function, averaged over environments, valid for both
the $T_1$ and $T_2$ relaxations. 
\SAVE{This is valid up to the 
averaging time scale, beyond which 
the monopole distribution should be
treated as ``annealed'' disorder.}
The key trick, allowing for the simple final result, 
is that (for either kind of relaxation) the combined 
relaxation function due to many flippers at sites
$\{ \RR_j \}$ is  simply a product $g^\comb_i(t)\equiv \prod_j g_{0i}(\RR;t)$.
(This follows immediately from the fact that both kinds of 
relaxation functions are products of independent random
variables depending on the respective flippers.)
\BRUTAL{
Now consider many independent flipping objects with density $n$.
(note the relaxation function factors into 
a product of relaxation functions from each independent source.)}
A (grand-canonical) ensemble is specified 
by setting the flipper occupation of each site to be 1
with probability $\nx$ or $0$ with probability $1-\nx$.~\cite{FN-poisson}
Defining $g_i(t)$ to be the ensemble average of $g_i^\comb(t)$,
I use conditional probability to obtain:
    \beq
    g_i(t) = \prod _\RR \Big( [1-\nx] + \nx g_{0i}(\RR;t)\Big)
     \approx e^{- \nx F_i(t)},
    \label{eq:g-n-F}
   \eeq
where, independent of $\nx$
   \bea
     F_i(t) &\equiv& \sum _\RR [1-g_{0i}(\RR;t)] 
     \approx \sum_\RR \big[ 1 - 
         e^{- \left(\frac{t}{T_{0i}}\right)^{\beta_{0i}} |\RR|^{-6}} \big] 
   \label{eq:F-sum-g} \\
        &\approx& c_6 (t/T_{0i})^{\beta_{0i}/2}
   \label{eq:F-sum-result}
    \eea
where I converted the sum in \eqref{eq:F-sum-g}
to an integral, using $\sum_\RR \to d^3\RR/\sqrt{2}$ 
in units of $1/\ap^3$, thus $c_6 = (4\pi/3) \int _0^\infty
(1-e^{-\xi^2})d(1/\xi) = (32\pi^3/9)^{1/2} \approx 10.5$.

We can qualitatively interpret  the result \eqr{eq:F-sum-result}
as follows.  More distant flippers contribute smaller fields
which take longer to decohere the probe spin; thus
in \eqr{eq:F-sum-g}, $g_{0i}(\RR;t)\approx 0$ at short
$|\RR|$ or 1 for large $|\RR|$. Hence the sum in \eqr{eq:F-sum-g}
roughly counts how many are in the first category; 
if you write $g_{01} \sim \exp(-R_*^6/|\RR|^6)$, then
$R^*$ is the effective radius within which we count
the sites so $F_i(t) \propto R_*^3$.

Incidentally it can be seen that if $\taux$ is temperature-independent, 
then the only temperature
dependence in \eqref{eq:g-n-F}  is via $\nx$; 
thus plots of $\ln g_i(t)$, taken at different temperatures,
ought to be identical, apart from an overall prefactor
which gives the temperature dependence of  $\nx(T)$.

\SAVE{More precisely I wrote
      $g_{0i}(\RR;t) = \exp[-c(\RRhat) (R_*/|\RR|)^6 ]$
where $c(\RRhat)$ includes the actual angular factors
in the dipole interaction (which I have fudged by using
the r.m.s. instead).
In place  of $c_6$ we have an integral proportional to
$C \equiv \int d^3\XX \big(1-\exp[-c(\XXhat)/|\XX|^6]\big)$.
But notice a regime I omitted in the paper:
if $R_*^3 \ll 1$, then 
$F(t) \approx C' R_*^6 $
where $C'\equiv \sum_\RR c(\RRhat)|\RR|^{-6}$.
}

The final result in all cases is a stretched exponential 
   \beq
      g_i(t) \propto \exp[-(t/T_i)^{\beta_i}]
   \label{eq:g-stretched-final}
   \eeq
where $\beta_i = \beta_{0i}/2$ and 
   \beq
     T_i^{-\beta_i} = c_6 \nx T_{0i}^{-\beta_i}.
   \label{eq:T_i-stretched-final}
   \eeq
If the ``flippers'' (such as monopoles) are thermally
excited, then the temperature dependence of $T_1$ and $T_2$
follows from $T_i \propto (\taux/\nx)^{1/\beta_i}$.
Thus if $\taux$ is temperature independent and
$\nx$ has an activation energy $\Delta$, the
activation energy for $T_i$ is $\Delta/\beta_i$.

\LATER{Cite Ref.~\cite{giblin-bramwell-mono-currents} for parameters.}

\secc{Experimental results and comparison to theory}

Unpublished work of Kitagawa, Takigawa, {\it et al}~\cite{kitagawa-lowT}
found stretched exponential forms for both relaxation functions,
   of form \eqr{eq:g-stretched-final}
The time scales are $T_1 \sim 1 $ s  
and $T_2 \sim 10^{-4}$ s at  0.5 K,
growing with decreasing temperature. Eventually they saturate
with $T_1 \sim 10^3$ s from 0.2 K downwards, while
$T_2\sim 10^{-4}$ s from 0.4 K downwards.
Both exponents $\beta_i$ decrease with temperature, 
starting with $\beta_i \approx 1$ 
(i.e. unstretched exponential) for $T>1$ K, it would appear
each exponent saturates to 1/2 at about the same temperature
that the corresponding $T_i$ saturates, 
so $\beta_1 < \beta_2$ at intermediate temperatures.
(As for the $^{47}$Ti NMR relaxations, the temperature was not taken far below 
$T^*$ so not much can be said about $T\to 0$ behavior, but at the temperatures
investigated, both $T_1$ and $T_2$ were an order of magnitude longer than they
were for $^{17}$O.)

The low-$T$ limiting behavior has
two fundamental contradictions with 
any diffusing monopole theory.
Rirst is the temperature dependence of $T_2$.
First, in Eq.~\eqref{eq:g-n-F}
the {\it only} $T$ dependence 
comes from the density of flippers $\nx$
(presumed in this picture to be monopole 
density $n$)
or conceivably the flip frequency $\taux^{-1}$.
In \eqref{eq:T_i-stretched-final}, this
\SAVE{There was an error in Kitagawa \& Takigawa seminar, 
where they assumed $1/T_i  \propto n$.}
implies $T_i$ must have an activated temperature dependence, which 
contradicts the observed saturation at low temperatures.  
(If $\taux$ were also activated, it would just {\it add} 
to the activated dependence.)

The second contradiction is that I found
$\beta_2=3/2$ in the monopole diffusion regime,
contradicting the experimental value $\beta_2 \approx 1/2$.
and indicating that $\beta_{02} \approx 1$.
But that holds only when the flipper is fixed and
flipping rapidly compared to the measurement time,
$\taux \ll T_2 \sim 10^{-4}{\rm s}.$

\SAVE{
It should be remembered that the cancellation of fields is not absolute in
the two-in/two-out spin states; there will be site-to-site spatial variations 
of the magnetic field, which at some temperature will exceed the typical
fields due to distant monopoles.  However, at low $T$, the monopole density is
suppressed by the Boltzmann factor, and the spin states 
(which have no other low-$T$ dynamics) should be entirely frozen, 
so the spin-echo method should fully cancel such site-to-site variations.}

One is forced by the data to assume a density of ``flippers''
that has negligible temperature dependence.
This can only be some kind of quenched disorder in the material, 
with a density perhaps $10^{-4}$--$10^{-3}$ per tetrahedron 
so as to dominate the monopole density at temperatures 
below $T^*$, where the temperature dependence of $T_1$ and $T_2$ levels off.  
From here on, let us accept that the flippers are fixed, 
giving $\beta_1=\beta_2=1/2$, and see what the experiment
tells us about them.

From the two relaxation times $T_1$ and $T_2$ we can
solve for the two unknown parameters $\taux$ and $\nx$.
Eliminating from Eqs.~\eqref{eq:T_0is}, we find
   \begin{subequations}
   \label{eq:n-taux-solve}
   \begin{eqnarray}
       \omega_0 \taux &\approx& \sqrt{2} (T_1/T_2)^{1/2}  \\
           n^{-1}      &\approx& \left(\frac{\omega_0}{\omega_d}\right)
                       {\Big(\omega_0 \sqrt{T_1 T_2}\Big)^{1/2}}
   \end{eqnarray}
   \end{subequations}
For Dy$_2$Ti$_2$O$_7$,
the two known parameters are $\omega_0/2\pi \approx  20$MHz,
and $\omegaD/\omega_0 \approx 1/20$.\\
Insertion of these and the experimental $T_i$'s into
Eqs.~\eqref{eq:n-taux-solve}
yields $\taux \approx 3.5 \times 10^{-5}\,{\rm s}$,
which is (barely) consistent with the condition $\taux \ll T_2$, 
and $\nx\approx 0.002$, still in units of $\apy^{-3}$.

What can the fixed flippers be?
The out-of-equilibrium, back-to-back bound monopole pairs are ruled out:
any fluctuations depend on a momentary  energy increase, but
the rate $\taux^{-1}$ would be thermally activated, contrary to observation.  
Dilution by nonmagnetic sites 
produces a very similar situation: pairs of unbalanced tetrahedra,
each of which is analogous to a midgap impurity in a semiconductor.
The minimum energy state adjacent to the nonmagnetic
site is a bound pair of half-monopoles, and the fluctuation rate
is again activated.

\BRUTAL{
The first candidate for a defect would be an isolated nonmagnetic site. 
That creates a tetrahedron on either side that behaves analogously to 
a midgap level in a seminconductor:
it can have either a two-in/one-out or a one-in/two-out state,
corresponding to half-monopole excitations of charge $\pm Q/2$.
The pair of tetrahedra have two degenerate ground states with
charges $(+Q/2,-Q/2)$ or $(-Q/2,+Q/2)$, due to the Coulombic interaction
between them.   
The activation energy to flip that spin (transiently) is still 
about $2\Jeff$ (not including the dipolar part), not so different
from the one controlling $\nmono$, so this cannot be the answer.
The situation is reminiscent of the back-to-back defects.}

\BRUTAL{The experimental
magnetic relaxation time constant is around $10^5$s (1 day)
at $T\approx 0.25$K~\cite{hiroi-mag-lowT}.
So at $T<T^*$ the equilibration time of the system might 
similarly be days: there might be a 
a certain non-equilibrium density of monopoles would be quenched in, 
the density of which (at shorter times) will not change with temperature.
A particular defect was identified of long-lived ``back-to-back'' pairs of
oppositely charged monopoles~\cite{castel-back2back},
which get pinned on neighboring 
diamond sites due to their attraction, but which cannot recombine since
the intervening spin is in the wrong direction.}

\SAVE{If we have two nonmagnetic sites in adjacent tetrahedra,
it is possible (if we consider only nearest-neighbor interactions)
for the shared spin to have zero local field, and thus be free
to flip with no energy cost.  However, in the monopole
language, this corresponds to configurations in which the 
respective half-monopoles are of the same sign; in the ground state,
they must have opposite signs.  (Taken together, the two diluted
sites make four defect tetrahedra in a chain, which must have
half-monopoles with alternating charges.)}

\SAVE{Nuclear spins typically fluctuate far too slowly.}

\BRUTAL{The best way out seems to be positing {\it interstitial} magnetic defects,
such as ``stuffed'' spin ice~\cite{stuffed,gardner-stuffed} 
in which a few magnetic atoms (Ho or in this case Dy) sit on the Ti 
(``B'') sites, which are  the centers of hexagons formed by the
magnetic Dy (``A'') sites.}

\BRUTAL{
Let us assume the ``stuffed'' Dy spins 
have an easy axis: this will be their local axis
(normal to the plane of that hexagon).  The spins in that hexagon
point mostly in that plane (in-plane components do not contribute 
to the effective local field on the Ising-like stuffed spin). More
importantly, due to the two-in/two-out rule on each tetrahedron,
there is a tendency for adjacent spins in the triangle to give
cancelling contributions to the out-of-plane field: I believe that,
in the spin-ice state the dipolar field at the stuffed site
due to these six closest spins will cancel at least half the time.
}
\OMIT{I ignore any perturbations due to the stuffed spin itself.}
The nearest out-of-plane spins are almost as close, but their easy axis
is oriented exactly so that the two terms in the dipole interaction
cancel.

\OMIT{The field of the six spins vanishes whenever all
six surrounding spins are pointing head-to-tail around the hexagon.
That should be the likeliest (though far from dominant) configuration
of those spins in a spin-ice state.  But there are other combinations
that also lead to zero.}

\SAVE{Presumably, the role of ``stuffed'' spins can be tested by growing samples with
deficient and excessive Dy content, favoring nonmagnetic dilutions or
magnetic interstitials, respectively; as I just explained, the former 
case does not seem to generate  low-$T$ spin-fluctuations but the 
latter case might.}

\SAVE{
The final possibility (as has been suggested somewhere) is that monopole 
diffusion is impeded by small effective potential barriers, induced
by the two-in/two-out background.}

\SAVE{Presumably a good way to distinguish the different scenarios of low
temperature relaxation is to repeat NMR experiments in nonzero field,
whenever possible.}

My best guess is that extra magnetic (Dy) spins 
appear on the nonmagnetic (Ti) site, as in the 
``stuffed spin ice''~\cite{stuffed,gardner-stuffed}
but much more dilutely.
(Note the Ti sites themselves form a pyrochlore lattice dual to the Dy
pyrochlore lattice.)
Indeed, it was proposed very recently~\cite{ross-stuffed}
that in the related material Yb$_2$Ti$_2$O$_7$, around 5\%
of the Ti sites are occupied by the magnetic Yb ion.

In order for the ``stuffed'' spins to fluctuate at such
with such a short time constant $\tau_0$, they must be
practically decoupled from the nearly frozen lattice
of regular Dy spins.  
This decoupling seems plausible when one
considers the location of the Ti sites, at the centers
of hexagons formed by Dy sites, and assumes the ``stuffed''
spin has an easy direction along its local three-fold axis.
First, the local field at the Ti site due to its six nearest-neighbor
pyrochlore spins cancels if they are all oriented in the same sense around the ring,
which should happen $\sim 12\%$ of the time.
Second, the dipole coupling to its second-nearest Dy spins 
happens to have an angular factor that exactly cancels.
\LATER{Check the cancellation.}
Indeed, it appears from Figure 1 of
Ref.~\cite{field-trip} that the typical local field at a Ti site
is $\sim 0.2$ Tesla or about $\sim 1/20$ of the maximum local field,
which occurs on the O(1) sites containing the probe nuclei.~\cite{FN-supertetrahedron}

\secc{Conclusion}

In conclusion, I have rederived the stretched-exponential form of
the NMR relaxation functions due to independent flipping spins
at random, distant positions, which might be either fixed impurities
(weakly coupled to any ordered or frozen spin background)
or else the spin-flips which induce the hopping of emergent monopole
defects in spin ice.  In particular, monopole hopping implies a novel
power of 3/2 in the stretched exponential for the $T_2$ relaxation,
in contrast to 1/2 for a fixed magnetic impurity.  

Experiment
at the lowest temperatures shows, for both $T_1$ and $T_2$ relaxations, 
a power tending to 1/2 in the stretched exponential and relaxation times  
tending to a constant, both of which are incompatible with the
monopole picture.  An analysis was presented that allows 
extraction of $\taux$ and $\nx$ from $T_1$ and $T_2$ with 
no bias as to the cause (except it depends crucially on 
both kinds of relaxation being due to the same fluctuations).
I suggested that dilute magnetic impurities ``stuffed''
on the Ti sites are responsible.  

It would be interesting to see if NMR relaxation
in the higher temperature regime around 0.5 K {\it can} be explained by monopoles. 
This may be more complex, though:
there may be no temperature range in which thermally excited monopoles are dense enough
to dominate over the ``stuffed'' impurities, and are at the same time dilute
enough for the small density approximations used here.

\secc{Acknowledgments}
I thank M. Takigawa, S. Bramwell, and S. Dunsiger for discussions, 
and M. Takigawa for providing Ref.~\cite{kitagawa-lowT}.
This work was supported by NSF grant DMR-1005466.

\OMIT{Switching between these two states requires 
separating a full monopole from one tetrahedron (the process $+Q/2 \to -Q/2 + Q$)
and moving it along a long loop to the opposite tetrahedron. Consequently
this process has a nonzero activation energy.}


\begin{thebibliography}{99}
\bibitem{FN-hop-anticorrelated} 
I have ignored a small anticorrelation between
successive hops, due to the dependence of available
hopping direction on spin configuration.
\SAVE{(Since I am not including the background spin
configuration as a degree of freedom, in this picture.)}
\bibitem{FN-nuc-moments}
The actual nuclei have larger moments but this does not
affect the results.
\bibitem{FN-NQR}
The behavior is practically the same in the case that $\omega_0$ 
is due to a uniaxial anisotropy acting on the nuclear spin 
(known as nuclear quadrupole resonance), as in the case of the $^{47}$Ti nuclei;
in any case $\hbar \omega_0$ is the level splitting.
\bibitem{FN-flips-displacement}
In fact, it is easy to see (using the Coulomb
formula for the monopole's field) that the accumulated
$\Delta \omega$ is a linear combination of components
of the accumulated displacement vector.
Thus the $\Delta \hvec$ jumps can add up without bound.
\end{thebibliography}
\end{document}